\documentstyle[11pt,newpasp,twoside]{article}
\def\edcomment#1{\iffalse\marginpar{\raggedright\sl#1\/}\else\relax\fi}
\reversemarginpar

\begin{document}
\title{New Southern Groups \altaffilmark {1}}
\altaffiltext {1}{Based on observations made under the Observat\'{o}rio
Nacional-ESO
agreement for the joint operation of the 1.52\,m ESO telescope and at the
Observat\'{o}rio do
Pico dos Dias, operated by MCT/Laborat\'{o}rio Nacional de
Astrof\'{\i}sica,  Brazil}

\author{R. de la Reza, L. da Silva, E. Jilinski}
\affil{Observat\'orio Nacional/MCT, Rua General Jos\'e Cristino 77,
S\~ao-Crist\'ov\~ao,
20921-400, RJ, Brazil}
\author{C. A. O. Torres, G. Quast}
\affil{Laborat\'{o}rio Nacional de Astrof\'{\i}sica/MCT, CP 21, 37504-364
Itajub\'{a},
MG, Brazil}

\begin{abstract}
Since the eighties we have begun, in Brazil, a search for Post-T Tauri
stars. Here we describe the main steps of this research that resulted in
the discovery of the nearby TW Hya and Horologium associations. A very
recent survey resulted in the detection of three different kinematical
groups:
1) the Great Austral Young Association, which is a very extended complex
region
involving the Horologium and Tucana associations, 2) a CrA
extend association and 3) a new group in Pisces.

\end{abstract}

\section{Introduction}

Post-T Tauri stars (PTTS) are young low mass stars which are in an
evolutionary stage between T Tauri stars (TTS) inserted in their forming
clouds with ages less than 10 Myrs and young active main sequence (MS)
stars
with ages near 100 Myrs as the Pleiades stars (see also Jensen in this
volume).
Our definition of PTTS also includes an ensemble of stellar properties,
that evolve
during that period of time. These properties keep, in general,
characteristic values
between those found in TTS and those of active main sequence stars. These
are:
X-ray emission, Lithium abundance, weak absorption or emission $H\alpha$
lines and
relative high stellar rotation. Together with this definition, we add a
supplementary
constraint used also as an strategy to discover these PTTS, and is that
they belong
to a moving group with similar spatial velocities. One consequence, for
instance,
of this evolutive
definition of PTTS, considering also the group concept, is that the
evolution of the
stellar disks can be studied.
We can see that very few PTTS exist such as TW Hya, Hen 600A and V4046 Sgr
still having their dusty disks, probably in their last stages of life.
The large majority of the
PTTS if they have not lost their accreting disks, they have probably
transformed them into rocky disks, preparing the formation of the first
planets.

Our search for PTTS started in the early eighties by studying TTS
isolated from clouds. The favorite targets were at that time V4046 Sgr
(HD 319139), AS216, AS218, FKSer and AB Dor (Quast et al. 1987). We learned
later that a similar list of stars including TW Hya had already been
proposed by Herbig (1978) as being good candidates for PTTS. Our
systematic observations of V4046 Sgr over the period of several years
enabled us to detect this star as the first Classical TTS being a double
line spectroscopic binary (de la Reza et al. 1986, Quast 1998, Quast
et al. 2000). The explanation of the isolated status of V4046 Sgr has
been a problem since. It is only now that we are probably finding a
solution by considering this star to be a member of the CrA extended
association (see Quast et al. in this volume).

At the end of the eighties we began a survey to detect new isolated TTS
based on IRAS sources properties. This survey called ``the Pico dos Dias
Survey''
(PDS) covered all the southern sky up to DEC $<$ +30 deg. As a test for
the methodology to be used in the PDS, we decided to search around TW
Hya. This initial operation was a success because the first members of
the future TW Hya association (TWA) were discovered.
That was the case of Hen 600AB and CoD-2988879 (de la Reza et al. 1989).
Already during the PDS, two more members were added to this group, these
were HD 98800 and CoD-337795 (Gregorio-Hetem et al. 1992).
Later, the knowledge of Hipparcos
parallaxes gave to TWA its nearby association status (Kastner et al. 1997).

During the PDS, we detected a number of other interesting astrophysical
objects, but no more isolated groups of PTTS. The reason became clear only
later; examples of Classical isolated TTS as TW Hya, Hen 600 A, HD 98800
and V4046
Sgr, contained in these stellar associations,
are very rare due to the rapid evolution of their dusty disk
stages, marking probably their end.

\section{The New Southern Hemisphere Survey}

It became clear that a much more efficient way to detect new PTTS
consists in using X-ray sources. Because of their relatively high velocity
rotation, PTTS result in efficient X-ray emitters. This turn to be evident
when
ROSAT X-ray measurements were used to detect additional members of TWA
(Jensen et al. 1998,
Webb et al. 1999, Sterzik et al. 1999, Jayawardhana et al. 1999 and
Zuckerman et al. 2001).
Later, at the end of the nineties, we began a new PTTS
survey using this time RASS X-ray sources instead of IRAS ones. However,
we operate with the same methodology as used in the discovery of TWA,
that is, searching around a pre-selected candidate field PTTS. In this way,
searching around PDS1 (Hen1) we discovered the nearby Horologium
association (Torres et al. 2000). At the same time, independently and
using a different approach based essentially on Hipparcos measurements, Zuckerman
\& Webb (2000) discovered another association in Tucana (TucA) having
similar properties of age $30-40$ Myrs and distance about 60 pc as HorA. It
is not only because of these similarities, but also, to the fact that both
associations are very close in the sky, that leave us to suppose, not
only
that both associations could be the same, but also  that an even a larger
association
involving HorA
and TucA could exist. A very large exploration area in the sky, going in
the direction of the South Pole was then made. We explored also a new
control area at
high galactic latitudes around the star BP Psc.
We call this new observational campaign a Survey for Associations Containing 
Young stars (SACY).

\section{Observations}

In the SACY we selected and observed all bright RASS sources that could be
associated with
TYCHO-2 or HIPPARCOS stars later than G0. Until now we examined    the
area:\\
\centerline{17:00 $< \alpha <$ 09:00 ~~and ~~\,~~~~~~~~~~$\delta
< -45\deg$}
\centerline{17:25 $< \alpha <$ 20:00 ~~and ~~$-45\deg < \delta < -23\deg$}
\centerline{00:00 $< \alpha <$ 06:00 ~~and ~~$-45\deg < \delta < -40\deg$}
\centerline{09:00 $< \alpha <$ 14:10 ~~and ~~~~~~~~~~~~~$\delta < -75\deg$}
This area engulfs the previous southern control region in Torres et al.
(2000).
The new control area around the possible CTT BP Psc was: \\
\centerline{22:40 $< \alpha <$ 01:00 ~~and ~~$-12\deg < \delta < +08\deg$}.

In all this area, which represent about 30\% of the Southern Hemisphere,
more than 400 stars were observed.
We obtained high resolution spectra for the selected candidates, with the
FEROS echelle spectrograph (Kaufer et al. 1999)
(resolution of 50000; spectral coverage of 5000\,\AA)
of the 1.52 m ESO telescope at La Silla
or  with the coud\'{e} spectrograph
(resolution of 9000; spectral coverage of 450\,\AA, centered at 6500\,\AA)
of  the 1.60\,m telescope of the Observat\'{o}rio do Pico dos Dias.
For some stars we obtained
radial velocities with the CORALIE at the Swiss Euler Telescope at ESO
(Queloz et al.
2000).

The FEROS spectra
enabled us to obtain good measurements of key lines for this type of
research as: Li, $H\alpha$, Ca II and Na D and to obtain reliable radial
and rotational velocities.

\section{New Groups}

The results of the SACY are very encouraging because three groups of PTTS
with relatively well
determined kinematical properties were found. The first one, and what we
called the Great Austral
Young Association (GAYA), is a very extended area containing the HorA and
TucA groups. GAYA has
nearly 44 stars with apparently very few binaries. The representative (U,
V, W) space velocities
for GAYA are:
(\mbox{U= $-9.8 \pm 1.2$}, \mbox{V = $-21.7 \pm 1.1$}, \mbox{W = $-2.0 \pm
2.2$}) .

Here, U, V and W in km/s, are positively measured in the directions of the
Galactic Center, Galactic Rotation
and North Galactic Pole
respectively. GAYA is estimated to have a size of about 60 pc, about half the members
of which are at the distance of about 50 pc and to be about 30 Myr old (see 
Torres et al. in this volume).

A second group,  independent of GAYA, was found. This one corresponds to
the CrA extended region.
The following
are their representative (U, V, W) space velocities  (\mbox{U= $-3.8 \pm
1.2$}, \mbox{V = $-14.3 \pm 1.7$}, \mbox{W = $-8.3 \pm 2.0$})
(see Quast et al. in this volume).

The results obtained for the third group around BP Psc were very
surprising. This is because the very few (four ``probable'' single stars
and one ``possible'' binary star) Li rich objects belong to the same moving
group with the following  (U, V, W)  space velocities,
obtained from the four ``probable'' stars: (\mbox{U= $-6.6 \pm 0.5$},
\mbox{V = $-0.6 \pm 0.3$}, \mbox{W = $-13.7 \pm 2.5$}),
which are very different from the values of the two above mentioned groups.

Unfortunately,
no Hipparcos distances are known for any of these Li rich objects.
Nevertheless, a minimum distance of about 75 pc can be inferred, on the one
hand because two of the Li-rich objects have their spectra
contaminated with interstellar Na D lines and on the other hand, from
the fact that five Li-poor nearby projected stars having Hipparcos
distances do not present this contamination. Assuming that these stars
with spectral types between G2 and K0 are on the MS, we
can obtain a reasonable distance of about 100 pc. In any case,
independently
of distances between 75 and 125 pc, the group maintains almost the same U
and V velocities. Other important properties for considering this group
as an association are the following: All these stars have the same Li
abundance of $log\epsilon (Li) = 3.2$ (where $log\epsilon (H) =12.00$) and
the same
X-ray ratio $(log Fx/log Fb) = -3.3$ (probably saturated). Also, rotational
vsini values are between 15 and 53 km/s, typical for PTTS, and visual
magnitudes are the same. Due to the absence of later type
stars we are unable for the moment to indicate an age for this
association, which we call ``Pisces association'' because all of its
members are in this constellation. These stars are localized in the
southern part of the known complex of translucid molecular clouds MBM53 and
MBM55.
However, we do not consider that there is any parental relation with
this cloud. A detailed presentation of this association can be found in
de la Reza et al. (2001).

\section {Conclusions}

Some results of the recent survey based on high-resolution spectra
Tycho-2 optical counterparts of X-ray RAAS sources in the Southern
Hemisphere consist in the detection of three independent moving groups
of PTTS, very probably placed at distances between 50 and 100 pc. Their
kinematical properties can maybe be understood if we place them into a
general distribution of space velocities in the solar neighborhood. If
we consider for instance, the general distribution of U and V velocities
of nearby stars (less than 100 pc) obtained by Skuljan et al. (1999), we
see
that stars are mainly concentrated in three branches called the
``Pleiades'', ``Middle'' and ``Sirius'' containing classical moving groups
of different
ages. The GAYA falls in the Pleiades branch and has similar, but not equal,
space
velocities to the Local Association (Pleiades supercluster ) being then
a fine structure of this supercluster. The same happens with TWA (Montes,
1999).
The extended CrA does not belong to the Local
Association and its velocities tend more to belong to the Middle branch.
The Pisces association falls precisely in the Middle branch. One future
possibility to understand the origin of these associations could be to
investigate the global effect (spiral arms?) that produce these
branches following the suggestion of Skuljan et al. (1999).

\acknowledgments

C. A. Torres thanks FAPEMIG, G. R. Quast CNPQ and R. de la Reza CAPES
for providing financial support.
This work was partially supported by a CNPQ grant to L. da Silva
(pr. 200580/97) and to C. Melo (proc. 200614/96-7).
We thank Dr M. Sterzik for instructive discussions
and M. Mayor for the use of the Swiss telescope.

\end{document}